%% file: Main.tex
\documentclass[12pt,final, draftclsnofoot, comsoc, onecolumn]{IEEEtran}

\usepackage[utf8]{inputenc}
\usepackage[english]{babel}
\usepackage{import}
\usepackage{nopageno}
\usepackage{graphicx,setspace,amssymb,graphics,dcolumn, array, multirow, float,epstopdf,paralist,lmodern,subcaption,balance,caption,subcaption,
	url,kantlipsum,multicol,enumerate,xfrac,soul}

\usepackage{amsmath}
\newtheorem{thrm}{Theorem}
\newtheorem{prop}[thrm]{Proposition}
\newtheorem{lem}[thrm]{Lemma} 
\newtheorem{corol}[thrm]{Corollary}
\newtheorem{remark}{Remark}
\newtheorem{mydef}{Definition}
\newtheorem{example}{Example}
 
\graphicspath{{./Figures/}}

\usepackage{filecontents}

\usepackage[english]{babel}
\usepackage[algo2e]{algorithm2e} 
\usepackage{algorithm}
\usepackage{algpseudocode}

\floatstyle{plain}
\newfloat{myalgo}{tbhp}{mya}

\newenvironment{Algorithm}[2][tbh]%
{\begin{myalgo}[#1]
		\centering
		\begin{minipage}{#2}
			\begin{algorithm}[H]}%
			{\end{algorithm}
		\end{minipage}
\end{myalgo}}

\begin{document}

\title{On the Average Locality of Locally Repairable Codes}
\author{Mostafa~Shahabinejad,~\IEEEmembership{Student~Member,~IEEE,}
~Majid~Khabbazian~\IEEEmembership{Senior~Member,~IEEE,}    
and~Masoud~Ardakani,~\IEEEmembership{Senior~Member,~IEEE}         
   
\thanks{The authors are with the Department of Electrical and Computer Engineering, University of Alberta, Edmonton, AB, Canada (Email: \{mshahabi, mkhabbazian, ardakani\}@ualberta.ca).} 
}
\maketitle
\renewcommand{\thefigure}{\arabic{figure}}

\begin{abstract}
A linear block code with dimension $k$, length $n$, and minimum distance $d$ is called a
locally repairable code (LRC) with locality $r$ if it can retrieve any coded
symbol by at most $r$ other coded symbols.
LRCs have been recently proposed and used in practice in distributed storage systems (DSSs) such as Windows Azure storage and Facebook HDFS-RAID. 
Theoretical bounds on the maximum locality of LRCs ($r$) have been established.  
The \textit{average} locality of an LRC ($\overline{r}$) directly affects the costly repair bandwidth, disk I/O, and number of nodes involved in the repair process of a missing data block.
There is a gap in the literature studying $\overline{r}$. 
In this paper, we establish a lower bound on $\overline{r}$ of arbitrary $(n,k,d)$ LRCs. 
Furthermore, we obtain a tight lower bound on $\overline{r}$ for a practical case where the code rate $(R=\frac{k}{n})$ is greater than $(1-\frac{1}{\sqrt{n}})^2$.
Finally, we design three classes of LRCs that achieve the obtained bounds on $\overline{r}$.
Comparing with the existing LRCs, our proposed codes improve the average locality without sacrificing such crucial parameters as the code rate or minimum distance.
\end{abstract}


\IEEEpeerreviewmaketitle

\section{Introduction}\label{sec:Introduction}
\input{Introduction.tex}

\section{Backgrounds and Definitions}\label{sec:backgrounds}
\input{Backgrounds.tex}

\section{Preliminary Results}\label{sec:prelimaries}
\input{preliminaries.tex}

\section{Lower Bounds on $\overline{r}$}\label{sec:genr_r_avg}
\input{genr_r_avg.tex}

\section{Achieving the Bounds: $\overline{r}$-Optimal LRCs}\label{sec:ProposedLRC}
\input{BoundAvgR.tex}

\section{Improvement of the LRC Used in Facebook HDFS-RAID}\label{sec:Comparison}
\input{ProposedLRC.tex}
\section{Conclusion and Future Work}
\input{Conclusion.tex}

\bibliographystyle{IEEEtran}
\bibliography{IEEEabrv,References}

\section{Proof of Theorem \ref{thm:avg_loc}}

\appendix[Proof of Theorem \ref{thm:avg_loc}]
\label{app:prfthm}
\input{AppAvgPrf.tex}

\end{document}

%% file: Introduction.tex
\subsection{Motivation}
Distributed storage systems (DSSs) provide a reliable, flexible, and cost-effective solution for digital data storage, and allow for anytime/anywhere access to one's data.  
Reliability and availability are achieved through redundancy.
For example, in the commonly used 3-replication method, three replicas of data are stored across distinct data nodes (DNs) \cite{Pyramid_codes,XORingElephants}.
Considering the rapid growth of data as well as costly maintenance of storage components in DSSs, 
the replication method is becoming unattractive due to its very large storage overhead.
Very recently, erasure codes have been used in DSSs to provide redundancy more efficiently \cite{Google_Availability,Pyramid_codes,XORingElephants}. 

An $(n,k,d)$ linear erasure code maps $k$ information symbols to $n$ encoded symbols $y_1$ to $y_n$ in a finite field $\mathbb{F}_q$, and has minimum distance $d$.
The upper bound on $d$ is obtained as $d\leq n-k+1$.
Erasure codes that achieve this bound are called maximum distance separable (MDS).
A downside of using conventional erasure codes in DSSs is their high-traffic data recovery \cite{XORingElephants}.
For example, in order for a scaler MDS code to recover a missing block, $k$ other encoded symbols have to be downloaded from other DNs. 
This problem that can be mitigated by using locally repairable codes (LRCs).
LRCs are designed such that recovering a lost/erased symbol requires access to a small number of other available symbols \cite{lrcDimakis,HomomDatta}.
The minimum number of symbols required to construct the $i$-th encoded symbol ($y_i$) defines its locality, and is denoted by Loc$(y_i)$.
The code locality ($r$) is defined as the maximum locality of its symbols.
Also, the code average locality ($\overline{r}$) is defined as the average of Loc$(y_i)$, i.e. $\overline{r}=\frac{1}{n}\sum_{i=1}^{n}\text{Loc}(y_i)$.
LRCs have been recently used in Facebook HDFS RAID \cite{XORingElephants} and Windows Azure Storage \cite{WAS}. 

\subsection{Related Work}
The connection between the code locality and other code parameters has been the subject of some recent studies \cite{lrcHuang,lrcDimakis}. 
In \cite{lrcHuang}, it is shown that $d\leq n-k-\lceil \frac{k}{r_{inf}}\rceil+2$, where $r_{inf}$ is the maximum locality of the information symbols.
In \cite{lrcDimakis}, it is verified that
\begin{equation}\label{eq:LRCbound}
d\leq n-k-\Big\lceil \frac{k}{r}\Big\rceil+2,
\end{equation}
where $r$ is the maximum locality of all symbols, i.e. $r=\underset{i \in [1,n]}{\max}\{\text{Loc}(y_i)\}$.
LRCs that achieve the bound in (\ref{eq:LRCbound}) are called optimal. 

In \cite{OptLRCDimakis}, optimal LRCs have been proposed over sufficiently large finite fields for the case that $(r + 1)|n$.
In \cite{shahabi_BLRC15_10_2014,goparaju_BLRC_2014,Zeh_Cyc_2015,shahabi_aclassBLRC} binary LRCs are proposed.   
In \cite{Cad_qBoundLRC_2013}, a bound on $k$ is obtained in terms of $n$, $k$, and $r$ as well as finite field order $q$.
In \cite{Silber_OptLRCAntCode_2015}, LRCs for small values of $r$ have been proposed some of which reach the bound in \cite{Cad_qBoundLRC_2013}.
In \cite{Wang_IntegerLRCBound_2015}, a tight bound on $d$ is obtained which improves the bound in (\ref{eq:LRCbound}) for some specific values of $n$, $k$, and $r$.

In some other schemes of LRCs, a missing block can be reconstructed by accessing multiple disjoint groups of other blocks, e.g. see \cite{Prak_optLRC_2012,Prakash_twoErasure_2014, Rawat_LocAndAvail_2014,Wang_MultErasure_2014} and references therein.
Other approaches in designing erasure codes for DSSs include reducing the repair bandwidth \cite{NCDimakis} and disk I/O \cite{rashmi_cake_2015}.

\subsection{Contribution of the Paper}
As shown in (\ref{eq:bound_r}), the bound in (\ref{eq:LRCbound}) is translated to a bound on the maximum locality $r$ in terms of $n$, $k$, and $d$.
The aim of this paper, however, is to $i)$ obtain lower bounds on $\overline{r}$ in terms of $n$, $k$, and $d$; and $ii)$ design LRCs that achieve the obtained bounds.   
In \cite{shahabi_ravg_2016}, as the first step towards establishing a bound on $\overline{r}$ and motivated by $(n,k,d)=(16,10,5)$ code used in Facebook HDFS-RAID \cite{XORingElephants}, 
we proved that the average locality $\overline{r}$ of any $(n,k)=(16,10)$ LRC with minimum distance $d=5$ is at least 3.875.
We also gave the construction of the LRC that achieves the bound $\overline{r}=3.875$.
In this paper, we generalize our work in \cite{shahabi_ravg_2016}. 
More specifically, we establish a lower bound on $\overline{r}$ of arbitrary $(n,k,d)$ LRCs (Theorem \ref{thm:genr_r_avg}). 
Furthermore, we obtain a tight lower bound on $\overline{r}$ for a practical case where $R=\frac{k}{n}>(1-\frac{1}{\sqrt{n}})^2$ (Theorem \ref{thm:avg_loc}).
Finally, we design three classes of LRCs that achieve the obtained bounds on $\overline{r}$ (Section \ref{sec:ProposedLRC}).
Note that the improvement achieved by our proposed LRCs comes without sacrificing such crucial coding parameters as rate $(\frac{k}{n})$ and minimum distance $d$.  
The following simple example shows the effectiveness of our solution.

\begin{example}\label{ex:8_4}
The existing solution: For an $(n,k,r)=(8,4,3)$ LRC, bound in (\ref{eq:LRCbound}) results in $d\leq 8-4-\lceil\frac{4}{3}\rceil+2=4$. In this case, since $(r+1)\mid n$, an optimal LRC with $d=4$ can be constructed whose Tanner graph is depicted in Fig. \ref{fig:1a}.
Here, Loc$(y_i)=r=\overline{r}=3$ for $i\in[1,8]$.

Our proposed solution: Fig. \ref{fig:1b} shows Tanner graph associated with our proposed LRC in this paper. 
For this LRC, Loc$(y_i)=2$ for $i\in [1,6]$ and  Loc$(y_i)=3$ otherwise.
Hence, $r=3$ and $\overline{r}=(6\times 2+2\times 6)/8=2.25$.
In other words, the average locality is improved by 25\% without changing $d$, and the code rate $k/n$. 
 
\begin{figure}
	\centering
	\begin{subfigure}[b]{0.27\textwidth}
		\includegraphics[width=\textwidth]{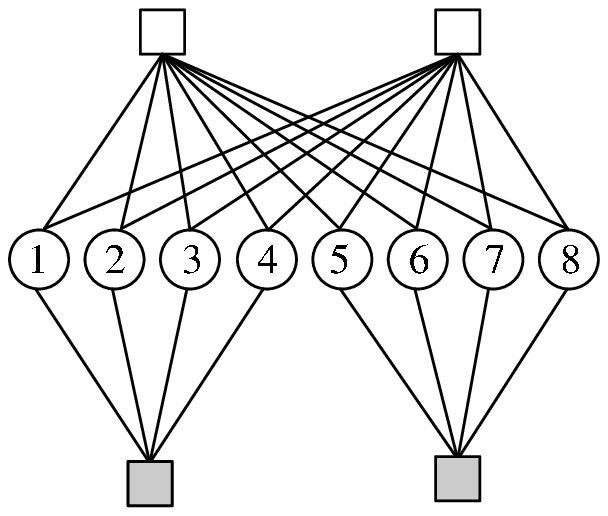}
		\caption{\footnotesize Tanner graph of an optimal $(n,k,d,r,\overline{r})=(8,4,4,3,3)$ LRC.}
		\label{fig:1a}
	\end{subfigure}
	\begin{subfigure}[b]{0.27\textwidth}
		\includegraphics[width=\textwidth]{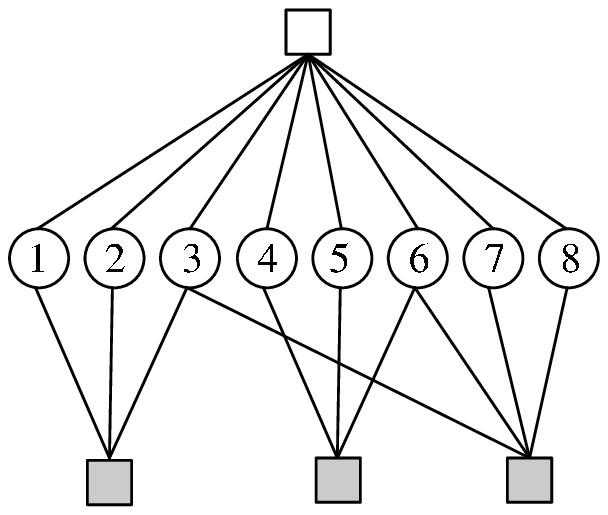}
		\caption{\footnotesize Tanner graph of our proposed $(n,k,d,r,\overline{r})=(8,4,4,3,2.25)$ LRC.}
		\label{fig:1b}
	\end{subfigure}
	\caption{ Tanner graph associated with two $(n,k,d)=(8,4,4)$ LRCs. Here, information symbols and parity symbols are depicted by white and gray circles. Linear combinations of symbols connected to each square is zero. 
	The average locality of the optimal LRC is improved by 25\% using our proposed LRC.}\label{fig:1}
\end{figure}
\end{example}

\subsection{Importance of The Results}
In practice, LRCs are of significant importance as they decrease the costly overhead associated with accessing nodes during a repair process \cite{lrcHuang}.
Also, the average locality of an LRC directly affects the repair bandwidth and disk I/O \cite{XORingElephants}. To show this, assume that $S_{DN}$ and $S_{Blk}$ represent size of a DN and size of a data block in terms of capacity unit, respectively.
Then, each DN can store $\frac{S_{DN}}{S_{Blk}}$ blocks of data. 
When a DN fails, 
on average, $\overline{r}\frac{S_{DN}}{S_{Blk}}$ blocks each of size $S_{Blk}$ have to be downloaded from active DNs in order to perform the  repair process.
In other words, the repair bandwidth is equal to $\overline{r}\frac{S_{DN}}{S_{Blk}}\times S_{Blk}=\overline{r}S_{DN}$.
Considering the large capacity of DNs (order of a few TB) as well as restricted repair bandwidth, even a small reduction in $\overline{r}$ can be of great importance \cite{XORingElephants}. 

Furthermore, mean-time to data-loss (MTTDL) associated with an LRC is inversely proportional to the
average locality of the code\footnote{MTTDL is a reliability metric defined as the average time taken such that a stripe of data cannot be recovered due to DN failures.} \cite{shahabi_aclassBLRC}.
In other words, for two LRCs with the same values of $(n,k,d)$ and different values of $\overline{r}$, the one with a smaller value of $\overline{r}$ is more reliable in terms of MTTDL.

In this work, by using a novel approach, we design LRCs with improved average locality.
In other words, the amount of costly repair bandwidth, disk I/O, number of participating nodes during repair process of a lost/erased block of data, and MTTDL are all improved by using our proposed coding schemes. 
Considering the recent interest in using efficient erasure codes in real-world data centers (e.g. in Google File System \cite{Google_Availability}, Windows Azure Storage \cite{WAS}, and Facebook HDFS-RAID \cite{XORingElephants}), our results can be of great importance from a practical point of view.
 
\textbf{Notations:} We show matrices and vectors by capital boldface letters and boldface letters, respectively. $\mathbb{F}_q$ stands for a finite field with cardinality $q$.
$\mathbf{I}_a$ and $(\cdot)^{T}$ represent an identity matrix of size $a$ and the matrix transpose operation, respectively.
For integers $a$ and $b$ with $b\geq a$, $[a,b]=\{a,a+1,\cdots,b\}$.
$\overline{\mathcal{A}}$ and $|\mathcal{A}|$ stand for the complement and cardinality of set $\mathcal{A}$, respectively.
For two sets $\mathcal{A}$ and $\mathcal{B}$, $\mathcal{A}\backslash\mathcal{B}$ stands for the relative complement of $\mathcal{B}$ in $\mathcal{A}$, i.e., $\mathcal{A}\backslash\mathcal{B}=\{a\in\mathcal{A}|a\notin\mathcal{B}\}$.
Finally, $\lfloor\cdot\rfloor$ and $\lceil\cdot\rceil$ represent the floor and ceiling operators, respectively.

%% file: Backgrounds.tex
\begin{mydef}(Systematic Linear Block Codes).
The generator matrix of an $(n,k,d)$ systematic linear block code $\mathcal{C}$ can be represented as $\mathbf{G} =[\mathbf{I}_{k}, \mathbf{P}] \in{\mathbb{F}_q^{k \times n}} $, where $\mathbf{P}\in{\mathbb{F}_q^{k \times (n-k)}}$. 
Hence, the encoded symbols can be obtained by $ \mathbf{y} = \mathbf{x}\mathbf{G}$, where 
$\mathbf{x}=[x_{1}, x_{2}, ..., x_{k}]\in{\mathbb{F}_q^{1 \times k}}$ 
and 
$\mathbf{y}=[y_{1}, y_{2}, ..., y_{n}]\in{\mathbb{F}_q^{1 \times n}}$ 
represent the information and encoded vectors, respectively. 
The parity check matrix of $\mathcal{C}$ is $\mathbf{H} =[-\mathbf{P}^{T}, \mathbf{I}_{n-k}] \in{\mathbb{F}_q^{(n-k) \times n}} $.
For any encoded vector $\mathbf{y}$ of $\mathcal{C}$, we have $\mathbf{y}\mathbf{H}^T=\mathbf{0}$.
Every $d-1$ columns of the parity check matrix of an $(n, k, d)$ linear block code are independent.
\end{mydef} 
 
\begin{mydef}(Minimum distance of code).
Assume that $\mathbf{u}$ and $\mathbf{v}$ denote two arbitrary distinct codewords of a code $\mathcal{C}$. 
Then, the minimum distance of $\mathcal{C}$ can be defined as $d =  \min\{d(\mathbf{u}, \mathbf{v})\}$, where $d(\mathbf{u}, \mathbf{v})$ is the Hamming distance between $\mathbf{u}$ and $\mathbf{v}$. 
In other words, $d$ is the minimum number of differences between any two codewords of $\mathcal{C}$.
For an $(n,k)$ code with the minimum distance of $d$, any $d-1$ symbol erasures can be reconstructed.
\end{mydef}

\begin{mydef}(Tanner graph).
	Tanner graph of an $(n,k)$ linear block code is a bipartite graph with two sets of vertices: a set of $n$ variable nodes and a set of $n-k$ check nodes. 
	The $i$-th variable node for $i\in [1,n]$ is adjacent with the $j$-th check node for $j\in [1,n-k]$ iff $\mathbf{h}_j(i)\neq 0$, where $\mathbf{h}_j$ represents the $j$-th row of the parity check $\mathbf{H}\in \mathbb{F}_q^{(n-k)\times k}$. 
	Therefore, variable nodes linked to a check node are linearly dependent.
\end{mydef}

In a DSS, linear block codes can be used to encode data as follows.
Assume that a stripe of data of size $L_S$ symbols in $\mathbb{F}_q$ is to be stored in a DSS.
The stripe is first broken into $k$ data blocks with size $l_S = \frac{L_S}{k}$ symbols. 
We denote by $x_{i,j}$ the $i$-th symbol of the $j$-th block. 
The coded vector is now obtained by $\mathbf{y}_i = [y_{i,1}, y_{i,2}, ..., y_{i,n}]=\mathbf{x}_i \mathbf{G}\in \mathbb{F}_q^{1\times n}$, where $\mathbf{x}_i=[x_{i,1}, x_{i,2}, ..., x_{i,k}]\in \mathbb{F}_q^{1\times k}$ with $ i\in[1,l_S]$. 
The $l_S$ encoded vectors $\mathbf{y}_i$'s are then stacked to constitute $\mathbf{Y}$ whose columns are $n$ encoded blocks stored in $n$ distinct DNs.
For the sake of simplicity, we assume $l_S=1$ from now on.

\begin{mydef}(Locality).
We denote the locality of an encoded symbol $y_i$ by Loc$(y_i)$ and define it as the minimum number of elements of $\mathcal{Y}\setminus \{y_i\}$ whose linear combination form $y_i$, where $\mathcal{Y}=\{y_1,...,y_{n}\}$. 
Therefore, there exists a subset of $\mathcal{Y}\setminus \{y_i\}$ indexed by $\mathcal{I}\subseteq [1,n]$ such that $|\mathcal{I}|=Loc(y_i)$ and
\begin{equation}\label{eq:adhset}
\forall~\mathbf{x},~~~\sum_{l\in\Theta}\alpha_ly_l=0,~~~\alpha_l \in \mathbb{F}_q, \alpha_l \neq 0,
\end{equation}
where Loc$(y_i)$ is the smallest possible integer and $\Theta=\{i\}\cup\mathcal{I}$. 
The average and maximum locality of the code are defined as $\overline{r}=\frac{\sum_{i=1}^{n}\text{Loc}(y_i)}{n}$ and $r=\underset{i \in [1,n]}{\max}\{\text{Loc}(y_i)\}$, respectively.
\end{mydef}

%

%% file: preliminaries.tex
In this section, we present definitions and lemmas required in discussing and proving our main results.
First, we show that there exists a Tanner graph, called locality Tanner graph, with at most $n-k$ CNs which determines the locality of all the encoded symbols.
Then, after defining and constructing local groups, we establish lower bounds on the average locality $\overline{r}$ in Section \ref{sec:genr_r_avg}.

A code does not have a unique Tanner graph representation.
In the following lemma, we define a representation of a Tanner graph, called locality Tanner graph, which reflects the locality of all the encoded symbols.

\begin{lem}\label{lem:MinLocTan}
A Tanner graph is called a locality Tanner graph, if there is a 
subset of CNs $\mathcal{P}_L\subseteq \mathcal{P}$, such that
\begin{inparaenum}[(i\upshape)]
	\item any $y_i\in\mathcal{Y}$ is connected to at least one CN in $\mathcal{P}_L$ with degree $\text{Loc}(y_i)+1$;
	\item check nodes in $\mathcal{P}_L$ are linearly independent (consequently,  $|\mathcal{P}_L|\leq n-k$).
\end{inparaenum}
CNs in $\mathcal{P}_L$ are called local CNs; others are called global CNs.
The set of VNs adjacent to a local CN is called a local group.
\end{lem}
\begin{IEEEproof}
	Our proof is by construction.
	First, a local group $\Psi_1$ with the minimum locality is selected from $\mathcal{Y}=\overline{\mathcal{U}_1}$, where $\mathcal{U}_1:=\{\}$.
	Then, a local group $\Psi_2$ with the minimum locality is selected from $\overline{\mathcal{U}_2}$, where $\mathcal{U}_2=\mathcal{U}_1\cup\Psi_1$.
	Similarly, a local group $\Psi_i$ with the minimum locality is selected from $\overline{\mathcal{U}_{i+1}}$, where $\mathcal{U}_{i+1}=\mathcal{U}_i\cup\Psi_i$.
	Note that each local group represents a CN.
	Since there are at most $n-k$ linearly independent equations of form (\ref{eq:adhset}), the algorithm terminates at most after $n-k$ local groups are selected, i.e. $|\mathcal{P}_L|\leq n-k$.
\end{IEEEproof}


\textit{Local group construction:}
By using a greedy algorithm, we partition the set of the encoded symbols $\mathcal{Y}=\{y_1,...,y_n\}$ into local groups $\mathcal{Y}_j$'s for $j\in[1,m]$ such that all elements of $\mathcal{Y}_j$ have the same locality $r_j$, where $r_1\leq r_2\leq \cdots \leq r_m$. 
The detailed procedure is described in the following algorithm.

\begin{Algorithm}[H]{8cm}
	\KwIn{$\mathcal{Y}=\{y_1,...,y_n\}$}
	\KwOut{$m,\mathcal{Y}_j,r_j$, for $j\in[1,m]$}
	\textbf{Initialization:} $j=1$, $\mathcal{U}_1=\{\}$\\
	\While{$|\mathcal{U}_j| < n$}{ 
		\begin{itemize}
			\vspace{1mm}
			\item Let $\Psi_j$ be the local group of an element in\\  $\arg\min\limits_{y_p\in~{\overline{\mathcal{U}}_j}} \{Loc(y_p)\}$,		where $\overline{\mathcal{U}}_j=\mathcal{Y}\setminus\mathcal{U}_j$
			\item Set $r_j=Loc(y)$, where~$y\in\Psi_j$
			\item $\mathcal{Y}_j\leftarrow\Psi_j\setminus\mathcal{U}_j$
			\item $\mathcal{U}_{j+1}\leftarrow\mathcal{U}_j\bigcup\mathcal{Y}_j$
			\item $j\leftarrow j+1$
		\end{itemize} }
		$m\leftarrow j$
		\caption{Local groups construction}\label{alg}
	\end{Algorithm}
	
	\begin{figure}[!t]
		\centering
		\includegraphics[width=0.5\textwidth, scale=0.6]{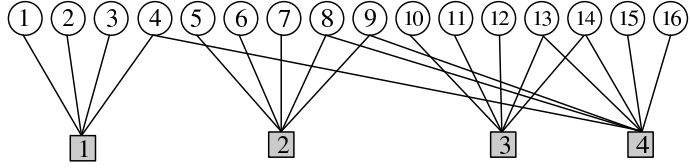}
		\caption{For the above locality Tanner graph, there are four local groups
			$\mathcal{Y}_1=\{y_1,y_2,y_3,y_4\}$, 
			$\mathcal{Y}_2=\{y_5,y_6,y_7,y_8,y_9\}$, 
			$\mathcal{Y}_3=\{y_{10},y_{11},y_{12},y_{13},y_{14}\}$, 
			$\mathcal{Y}_5=\{y_{15},y_{16}\}$. Also, $r=6$ and $\overline{r}=(4\times 3+10\times 4+2\times 6)/16=4$.
			The global CNs are not shown in this figure.}
		\label{fig:MinLocTannerA}
	\end{figure}
	\begin{mydef} \label{def:NonOverLCNs}(Non-overlapping local groups).
		We say that local groups one to $a$ are non-overlapping if intersection between any two of $\mathcal{Y}_1$ to $\mathcal{Y}_a$ is empty, i.e. $\mathcal{Y}_s\bigcap\mathcal{Y}_w=\emptyset$ for any distinct $s$ and $w$ in $[1,a]$.
	\end{mydef}

	\begin{example}
		Fig.~\ref{fig:MinLocTannerA} 
		shows a locality Tanner graph with $n=16$ VNs and $m=4$ local CNs.
		In this figure, global CNs are not shown.
		This locality Tanner graph has four local groups $\mathcal{Y}_1$ to $\mathcal{Y}_4$.  
	\end{example}

	
	The following Lemma shows how Tanner graph of an $(n,k)$ linear block code is related to the minimum distance of the code ($d$).
	
	\begin{lem}\label{lem:14_4}
		Over a sufficiently large finite field, a necessary and sufficient condition for an $(n,k)$ linear block code $\mathcal{C}$ with Tanner graph $\mathcal{T}$ to have minimum distance $d$ is that
		every $\gamma$ CNs of $\mathcal{T}$ cover at least $\gamma+k$ VNs for every $\gamma \in[J,~n-k]$, where $J:=n-k-d+2$.
	\end{lem}
	
	\begin{IEEEproof}
		\textit{Necessary condition (assumption: the minimum distance is $d$):}
		By contradiction, assume that there is a subset of CNs of $\mathcal{T}$ with cardinality $\gamma$ covering at most $\gamma+k-1$ VNs. Also, assume that the $n-(\gamma+k-1)\in [1,d-1]$ VNs not covered by the mentioned $\gamma$ CNs are erased. 
		Observe that among all $\gamma+k-1$ VNs covered by the mentioned $\gamma$ CNs, there are at most $(\gamma+k-1)-\gamma=k-1$ independent VNs.
		This implies that $\mathcal{C}$ cannot recover $k$ information symbols for the assumed $n-(\gamma+k-1)\in [1,d-1]$ erasures.
		This contradicts the fact that an erasure code with minimum distance $d$ can recover $k$ information symbols for any up to $d-1$ symbol erasures. 
		
		\textit{Sufficient condition
			(assumption: every $\gamma$ CNs of $\mathcal{T}$ cover at least $(\gamma+k)$ VNs for every $\gamma \in[J,~n-k]$):}
		An erasure code with minimum distance $d$ can recover any up to $d-1$ erasures.
		Let us define $\beta$ as $\beta:=(n-k)-\gamma+1$.
		Since $\gamma\in [n-k-d+2,~n-k]$, we have $\beta\in[1,d-1]$.
		Now, we show that any $\beta$ VNs are covered by at least $\beta$ CNs. 
		Over a sufficiently large finite field $\mathbb{F}_q$, this implies that any $\beta\in[1,d-1]$ erasures can be recovered using $\beta$ independent equations associated with their corresponding CNs.
		By contradiction, assume that there is a subset of VNs with cardinality $\beta$  covered by at most $\beta-1$ VNs.
		Then, the remaining $(n-k)-(\beta-1)=\gamma$ CNs cover at most $n-\beta=k+\gamma-1$ VNs. This contradicts the assumption that any $\gamma$ CNs cover at least $k+\gamma$ VNs.  
	\end{IEEEproof}
	\begin{remark}
		As we will see, all LRCs discussed in this paper have the following two properties: 
		$i)$ each local CN is connected to at least one VN which is not connected to other local CNs, and 
		$ii)$ each global CN is linked to all $n$ VNs.
		When these conditions are satisfied, Lemma \ref{lem:14_4} results in the following corollary which allows us to verify that the minimum distance of a code is $d$ by showing that every $J$ local CNs cover at least $J+k$ VNs. 
	\end{remark}

	\begin{corol}\label{cor:d_cndn}
		For an LRC with Tanner graph $\mathcal{T}$, assume that i) each local CN is connected to at least one VN which is not connected to other local CNs, and ii) each global CN is linked to all $n$ VNs.
		Then, the sufficient condition of Lemma \ref{lem:14_4} is equivalent to the following one.
		If every $J$ local CNs cover $J+k$ VNs, then the minimum distance of the code is $d$. 
	\end{corol}
	\begin{IEEEproof}
		Without loss of generality, consider the first $J$ local CNs that cover at least $J+k$ VNs.
		By adding $\gamma-J$ arbitrary local CNs to these $J$ local CNs, at least $\gamma-J$ more VNs are covered, where $\gamma\in[J,n-k]$. 
		This is true because each local CN is connected to at least one VN which is not connected to other local CNs. 
		Also, each global CN is connected to all VNs. 
		Hence, every $\gamma$ CNs cover at least $\gamma+J$ VNs.
	\end{IEEEproof}
	As we will see later, Lemma \ref{lem:14_4} is a strong tool which will help us obtain a bound on $\overline{r}$. 
	In fact, the famous bound of (\ref{eq:LRCbound}) on the maximum locality also can easily be obtained by Lemma \ref{lem:14_4}. 
	
	\begin{remark}
		For an $(n,k,d,r)$ LRC, assume that the number of local CNs is $m$, where $m\in\{2,\cdots,J,J+1,\cdots,n-k\}$.
		Now, we consider two cases as follows.\\
		Case (i) $m\in[J,n]$: In this case, by Lemma \ref{lem:14_4}, every $\gamma$ local CNs cover at least $\gamma+k$ VNs. We have
		\begin{equation}\label{eq:Cb}		
		r+1 \geq \frac{\gamma+k}{\gamma} \geq \frac{J+k}{J}, \quad m\in[J,n-k ].
		\end{equation}
		Case (ii) $m\in[2,J-1]$: In this case, all the $m$ local CNs cover all the $n$ VNs. We have
		\begin{equation}\label{eq:Ca}
		r+1 \geq \frac{n}{m} \geq \frac{n}{J-1}> \frac{J+k}{J}, \quad m\in[2,J-1],
		\end{equation}
		where the last equality holds because $n\geq J+k=n-(d-2)$.
		From (\ref{eq:Cb}), we have 
		\begin{equation*}
		r\geq \frac{k}{J} \Rightarrow n-k-d+2\geq \frac{k}{r} \Rightarrow d\leq n-k-\frac{k}{r}+2
		\end{equation*}
		Since $d$ is an integer, $d\leq n-k-\Big\lceil\frac{k}{r}\Big\rceil+2$.
	\end{remark}
	
	The following Lemma is used to partition a subset of the encoded symbols, say $\mathcal{A}$, with cardinality $A$ into $\zeta$ non-overlapping local groups such that the average locality of the encoded symbols in $\mathcal{A}$ is minimized.
	
	\begin{lem}\label{lem:OptSqrt}
		Let $z_j$, $j\in [1,\zeta]$ be integers, and $\sum_{j=1}^\zeta z_j=A$.
		Then,
		\begin{equation*}\label{eq:OptSqrt}
		\sum_{j=1}^\zeta z_j^2 \geq (\zeta-a) \Big\lfloor\frac{A}{\zeta}\Big\rfloor^2+a\Big\lceil\frac{A}{\zeta}\Big\rceil^2,
		\end{equation*}   
		where $a=A+\zeta-\zeta\lceil\frac{A}{\zeta}\rceil$.
	\end{lem}
	
	\begin{IEEEproof}
		Subject to $\sum_{j=1}^\zeta z_j=A$ and integer $z_j$, the sum $\sum_{j=1}^\zeta z_j^2$ is minimized if $|z_l-z_m|\leq 1$ for every pair $z_l$ and $z_m$. 
		Because otherwise if $z_l>z_m+1$, then $\sum_{j=1}^\zeta z_j^2$ can be reduced by setting $z_l$ to $z_l-1$ and $z_m$ to $z_m+1$.
		This is true since 
		\begin{equation*}
		z_l>z_m+1 \Leftrightarrow z_l^2+z_m^2\geq (z_l-1)^2+(z_m+1)^2.
		\end{equation*}
		Consequently, $\sum_{j=1}^\zeta z_j^2$ is minimized if for every $j$, $z_j=z$ or $z_j=z-1$ for some integer $z$. 
		The number $z$ is unique and it is a function of $A$ and $\zeta$.
		Now, assume that among $\zeta$ integers $z_j$, $a$ integers are $z$ and the rest $\zeta-a$ integers are $z-1$. 
		Therefore, $az+(\zeta-a)(z-1)=A$; equivalently, $a=A+\zeta(1-z)$.
		Hence, $\lceil\frac{a}{\zeta}\rceil=\lceil\frac{A}{\zeta}\rceil+1-z$. Therefore, $z=\lceil\frac{A}{\zeta}\rceil$ because $a\leq \zeta$.
	\end{IEEEproof}


%% file: genr_r_avg.tex
First, in Section \ref{subsec:genr_r_avg}, we derive a lower bound on $\overline{r}$ that holds for any $(n,k,d)$ LRCs.
We compare this bound to the one on  maximum locality $r$.
In Section \ref{subsec:spec_bound}, we obtain a tight lower bound on $\overline{r}$ for $(n,k,d)$ LRCs which satisfy the following constraint on the code rate $R> \Big(1-\frac{1}{\sqrt{n}}\Big)^2$.
In section \ref{sec:ProposedLRC}, we present three classes of LRCs that achieve the obtained bounds presented in this section.

\subsection{A Lower Bound on $\overline{r}$ for Arbitrary $(n,k,d)$ LRC}\label{subsec:genr_r_avg}


First, let us reform the bound in (\ref{eq:LRCbound}) to obtain a lower bound on $r$.
From (\ref{eq:LRCbound}), we have
\begin{multline*}
d\leq n-k-\Big\lceil\frac{k}{r}\Big\rceil+2 \leq n-k-\frac{k}{r} +2 \Rightarrow 
r\geq \frac{k}{n-k-d+2} =\frac{k}{J}.
\end{multline*}
Since $r$ is an integer, the lower bound of the maximum locality $(r)$ can be presented as
	\begin{equation}\label{eq:bound_r}
	r\geq \Big\lceil\frac{k}{J}\Big\rceil.
	\end{equation}
 
\begin{thrm}\label{thm:genr_r_avg}
	For any $(n,k,d)$ LRC with $J=n-k-d+2$, the average locality of all symbols ($\overline{r}$) is bounded as follows
\begin{equation}\label{eq:gen_thm}
\overline{r}\geq\Big\lceil\frac{k}{J}\Big\rceil\Big(1-\frac{J\Big\lceil\frac{k}{J}\Big\rceil-k}{n}\Big).
\end{equation} 
\end{thrm}
\begin{IEEEproof}
First, we construct local groups using Algorithm \ref{alg}.
For the code to have the minimum distance of $d$, by Lemma \ref{lem:14_4}, the first $J$ local groups must cover $n-\theta$ VNs for some integer $\theta\in[0,d-2]$.
\[\sum_{i=1}^{n-\theta}\text{Loc}(y_i)=\sum_{j=1}^{J}|\mathcal{Y}_j|r_j\geq\sum_{j=1}^{J}|\mathcal{Y}_j|(|\mathcal{Y}_j|-1),\]
where $\sum_{j=1}^{J}|\mathcal{Y}_j|=n-\theta$.
Observe that the minimum of $\sum_{j=1}^{J}|\mathcal{Y}_j|(|\mathcal{Y}_j|-1)$ is obtained when 
$\sum_{j=1}^{J}|\mathcal{Y}_j|$ is minimized, i.e. when $\sum_{j=1}^{J}|\mathcal{Y}_j|=n-(d-2)=J+k.$
In this case, by lemma \ref{lem:OptSqrt}, 
there are $(J-a_{d-2})$ local groups with cardinality $(\lfloor\frac{k}{J}\rfloor+1)$ and locality $\lfloor\frac{k}{J}\rfloor$; and
$a_{d-2}$ local groups with cardinality $(\lceil\frac{k}{J}\rceil+1)$ and locality $\lceil\frac{k}{J}\rceil$,
where $a_{d-2}=k+J-J\lceil\frac{k}{J}\rceil$.

Note that according to Algorithm \ref{alg}, all VNs not in the first $J$ local groups have locality greater than or equal to the maximum locality of the first $J$ local groups.
In order to obtain a lower bound on $\overline{r}$,
we assume that all the remaining $d-2$ VNs not in the first $J$ local groups 
have locality equal to $\lceil \frac{k}{J}\rceil$, because for any $\theta$, we have$\max\limits_{j\in[1,J]}r_j\geq \lceil\frac{k}{J}\rceil$.
Hence, the minimum of $\overline{r}$ is achieved if there are 
$m_1:=(J-a)(\lfloor\frac{k}{J}\rfloor+1)$ VNs with locality $\lfloor\frac{k}{J}\rfloor$
and $n-m_1$ VNs with locality $\lceil\frac{k}{J}\rceil$.
In other words, 
\begin{equation}\label{eq:r_avg_tmp}
n\overline{r}\geq m_1\Big\lfloor\frac{k}{J}\Big\rfloor+{(n-m_1)}\Big\lceil\frac{k}{J}\Big\rceil.
\end{equation}
If $J\mid k$, then by replacing $a=J$ and $m_1=0$ in (\ref{eq:r_avg_tmp}), we have $\overline{r}\geq\lceil\frac{k}{J}\rceil$.
If $J\nmid k$, then by replacing  $m_1=(J\lceil\frac{k}{J}\rceil-k)\lceil\frac{k}{J}\rceil$ in (\ref{eq:r_avg_tmp}), we have $\overline{r}\geq\lceil\frac{k}{J}\rceil(1-\frac{m_1}{n})$.
\end{IEEEproof}

%
\begin{remark}
Considering (\ref{eq:r_avg_tmp}), the bound on $\overline{r}$ can be represented as follows
	\begin{equation}\label{eq:gen_thmm_1r_1}
\bar{r}\geq \alpha \Big\lceil \frac{k}{J} \Big\rceil + (1-  \alpha) \Big\lfloor \frac{k}{J} \Big\rfloor,
	\end{equation} 
where $\alpha=\frac{1}{n}(J\lceil\frac{k}{J}\rceil-k)(\lfloor\frac{k}{J}\rfloor+1)$.
\end{remark}

\begin{remark}
	The gap between the bounds on $r$ and $\overline{r}$, represented in (\ref{eq:bound_r}) and (\ref{eq:gen_thm}) respectively, is maximized when $J\mid(k-1)$.
	To see this, observe that by subtracting the right hand side of (\ref{eq:gen_thm}) from that of (\ref{eq:bound_r}), we have
	\begin{equation}\label{eq:diff}
	\frac{1}{n}\Big\lceil\frac{k}{J}\Big\rceil\Big(J\Big\lceil\frac{k}{J}\Big\rceil-k\Big)=
	\left\{ 
	\begin{array}{l l}
	\frac{1}{n}(J-\beta)\lceil\frac{k}{J}\rceil , & \quad J\nmid k\\
	0, & \quad J\mid k
	\end{array}\right.
	\end{equation}
	where $\beta=k\mod J$, and $\beta\neq 0$. 
	Hence, (\ref{eq:diff}) achieves its maximum value when $\beta=1$, equivalently, when $J\mid(k-1)$.
	In Section \ref{subsec:2nd}, we show that for sufficiently large values of $d$, the bound in Theorem \ref{thm:genr_r_avg} is achievable when $J\mid(k-1)$.
\end{remark}

\subsection{A Tight Lower Bound on $\overline{r}$ of LRCs with $R> \Big(1-\frac{1}{\sqrt{n}}\Big)^2$}\label{subsec:spec_bound}

In the previous section, in order to obtain a lower bound on $\overline{r}$, we assumed that the first $J$ local groups of $(n,k,d)$ LRCs cover $J+k$ VNs. 
Also, we assumed that locality of the remaining $n-(J+k)=d-2$ is equal to maximum locality of the first $J+k$ VNs.
The obtained bound in Theorem \ref{thm:genr_r_avg} is not always tight.
In this section, we obtain a tight lower bound on $\overline{r}$ assuming that 
\begin{equation}\label{eq:cond_R}
R=\frac{k}{n}> \Big(1-\frac{1}{\sqrt{n}}\Big)^2.
\end{equation}

It is notable that linear block codes currently in use---e.g. the $(16,10,5)$ code used in Facebook HDFS-RAID, the $(16,12,4)$ code used in Windows Azure Storage, and the $(9,6,4)$ code in Google File System---satisfy the condition presented in (\ref{eq:cond_R}).

\begin{figure*}[t!]
	\centering
	\includegraphics[width=0.75\textwidth, scale=1]{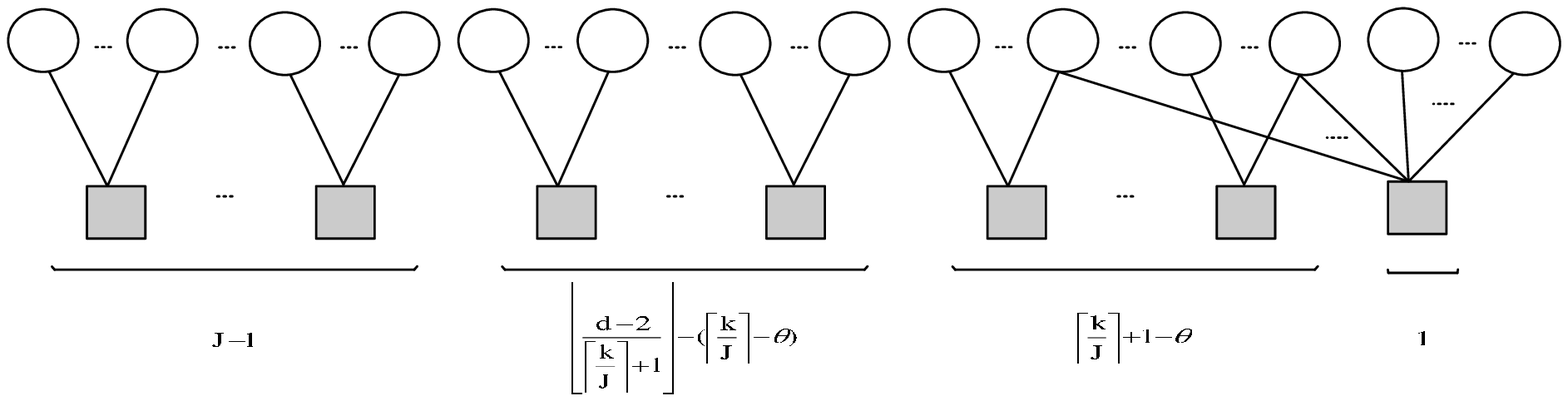}
	\caption{Locality Tanner graph of an $(n,k,d)$ LRC with $J\mid(k-1)$ and $\Big\lfloor\frac{d-2}{\lceil\frac{k}{J}\rceil+1}\Big\rfloor\geq \Big\lceil\frac{k}{J}\Big\rceil-\theta$. There are $(J-1)$ non-overlapping local groups with cardinality $(\lfloor\frac{k}{J}\rfloor+1)$; $\Big\lfloor\frac{d-2}{\lceil\frac{k}{J}\rceil+1}\Big\rfloor-(\Big\lceil\frac{k}{J}\Big\rceil-\theta)$ non-overlapping local groups with cardinality $(\Big\lceil\frac{k}{J}\Big\rceil+1)$; 
	$(\Big\lceil\frac{k}{J}\Big\rceil+1-\theta)$ overlapping local groups with cardinality $(\Big\lceil\frac{k}{J}\Big\rceil+1)$; and one overlapping local group which has $(\Big\lceil\frac{k}{J}\Big\rceil+1-\theta)$ VNs in common with $(\Big\lceil\frac{k}{J}\Big\rceil+1-\theta)$ distinct local groups.}
	\label{fig:ach_ravg_bound}
\end{figure*}

\begin{thrm}\label{thm:avg_loc}
	For any $(n,k,d)$ LRC with $R=\frac{k}{n}> \big(1-\frac{1}{\sqrt{n}}\big)^2$, the average locality of all symbols ($\overline{r}$) is bounded as follows
	\begin{equation}\label{eq:thm}
	\small
	\overline{r}\geq 
	\frac{
		\underset{\theta\in[0,d-2]}\min	
		\Bigg\{ 
		 (J-a_\theta)\Big\lfloor\frac{n-\theta}{J}\Big\rfloor^2
		 +a_\theta\Big\lceil\frac{n-\theta}{J}\Big\rceil^2
		 +(n-dJ+2J)\theta
		\Bigg\}
		}{n}
	-1,
	\end{equation} 
	where $J=n-k-d+2$ and
	$a_\theta=n-\theta+J-J\Big\lceil\frac{n-\theta}{J}\Big\rceil$.
\end{thrm}
\begin{IEEEproof}
	Here, we present the sketch of the proof.
	Please find the detailed proof in Appendix \ref{app:prfthm}.

	To begin with, we construct $m$ local groups using Algorithm \ref{alg}.
	Note that the maximum locality of an $(n,k)$ linear block code is $k$, hence, $m> 1$.  
	Furthermore, by Lemma~\ref{lem:MinLocTan}, $m\leq n-k$.
	Therefore,
	\begin{equation}\label{eq:r_avg_general}
	n~\overline{r} = \sum_{i=1}^n\text{Loc}(y_i)= \sum_{j=1}^{m}|\mathcal{Y}_j|r_j,~~m \in [2,n-k]. 
	\end{equation}
	
	Now, we consider two cases.
	As the first case, we assume that the total number of local groups is less than or equal to $J$, i.e. $m\in[2,J]$. 
	In this case, we verify that the minimum average locality is achieved when $m=J$.
	
	As the second case, we assume that the total number of local groups is greater than $J$, thus  $m\in [J+1, n-k]$. In this case, we verify that
	the minimum average locality is obtained when there are $m=J+1$ local groups.
	By Lemma \ref{lem:14_4}, in order to achieve minimum distance $d$, the first $J$ local groups must cover at least $J+k=n-(d-2)$ VNs and at most $n-1$ VNs
	\footnote{Note that the case that the first $J$ local groups cover $n$ VNs is equivalent to the first case with $m=J$. This case is considered in Theorem \ref{thm:avg_loc} by setting $\theta=0$.}.
	Among all possible $J+1$ local groups,
	we assume that the first $J$ local groups and the last local group cover $n-\theta$ and $\theta$ VNs, respectively, where $\theta\in[1,d-2]$. 
	Then, we obtain a lower bound on $\overline{r}$ according to this assumption.
	Finally, the lower bound on $\overline{r}$ is obtained by taking the minimum of two bounds associated with the considered cases.
\end{IEEEproof}

In Section \ref{subsec:3rd}, we design a class of LRCs that achieve the bound on $\overline{r}$ obtained in Theorem~\ref{thm:avg_loc}.

%% file: BoundAvgR.tex
In this section, we design LRCs that achieve the lower bounds on $\overline{r}$ obtained in Theorems \ref{thm:genr_r_avg} and \ref{thm:avg_loc}.
We call such codes, $\overline{r}$-optimal LRCs.
In Sections \ref{subsec:1st} and \ref{subsec:2nd}, we construct two classes of $\overline{r}$-optimal LRCs that achieve the general bound in Theorem \ref{thm:genr_r_avg}.
Furthermore, in Section \ref{subsec:3rd}, we construct a class of LRCs that achieve the bound in Theorem \ref{thm:avg_loc}.

\begin{figure*}[t!]
	\centering
	\includegraphics[width=0.75\textwidth, scale=1]{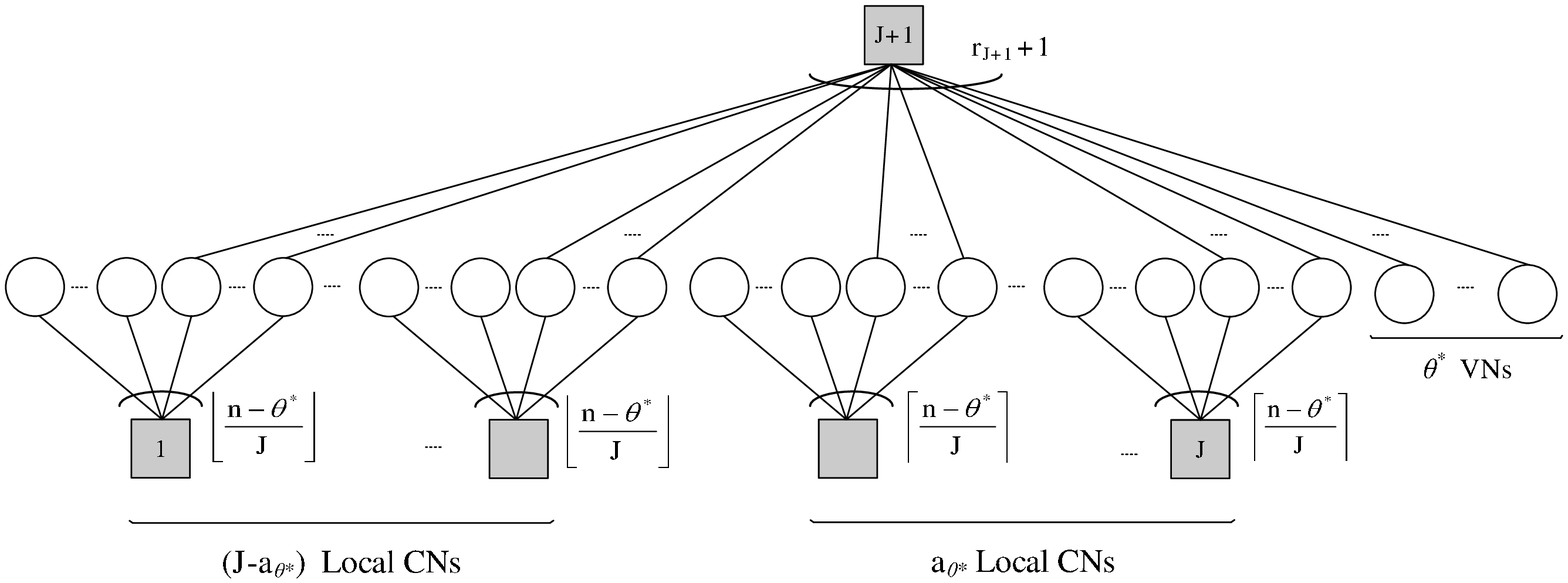}
	\caption{Locality Tanner graph of $\mathcal{C}_{\overline{r}LRC_3}$ with $J+1$ local CNs. 
	}
	\label{fig:r_avg_prop}
	\vspace{-2mm}
\end{figure*}
\subsection{The First Class of $\overline{r}$-Optimal LRCs}\label{subsec:1st}
Suppose the parameters $n$, $k$ and $d$ are such that $(\lceil\frac{k}{J}\rceil+1)\mid (d-2)$.
Then, the first class of our proposed $\overline{r}$-optimal LRCs ($\mathcal{C}_{\overline{r}LRC_1}$) achieving the bound in Theorem \ref{thm:genr_r_avg} can be constructed as follows.
The first $k+J$ VNs are partitioned into $J\lceil\frac{k}{J}\rceil-k$ and $J+k-J\lceil\frac{k}{J}\rceil$ local groups with cardinality $\lfloor\frac{k}{J}\rfloor+1$ and $\lceil\frac{k}{J}\rceil+1$, respectively.
Then, $n-(J+k)=d-2$ VNs not covered by the first $J$ local CNs are partitioned into  $\frac{d-2}{\lceil\frac{k}{J}\rceil+1}$ local groups with cardinality $\lceil\frac{k}{J}\rceil+1$.
Hence, there are exactly $\frac{m_1}{\lfloor\frac{k}{J}\rfloor+1}=J\lceil\frac{k}{J}\rceil-k$ 
and
$\frac{n-m_1}{\lceil\frac{k}{J}\rceil+1}$ 
local groups with cardinality 
$\lfloor\frac{k}{J}\rfloor+1$ and $\lceil\frac{k}{J}\rceil+1$, respectively.
Observe that for this class of $\overline{r}$-optimal LRCs, the constraint on $d$ expressed in Corollary \ref{cor:d_cndn} is satisfied and the bound in Theorem \ref{thm:genr_r_avg} is achieved.

\subsection{The Second Class of $\overline{r}$-Optimal LRCs}\label{subsec:2nd}
Suppose the parameters $n$, $k$ and $d$ are such that $J\mid (k-1)$ and
\begin{equation}\label{eq:cond_sec}
\bigg\lfloor\frac{d-2}{\lceil\frac{k}{J}\rceil+1}\bigg\rfloor\geq \Big\lceil\frac{k}{J}\Big\rceil-\theta,
\end{equation}
where 
\begin{equation}\label{eq:gen_theta}
\theta=(d-2)\mod (\Big\lceil\frac{k}{J}\Big\rceil+1).
\end{equation}
Fig. \ref{fig:ach_ravg_bound} shows the Tanner graph corresponding to our second class of $\overline{r}$-optimal LRCs ($\mathcal{C}_{\overline{r}LRC_2}$).
In this graph, all the local groups, except $\lceil\frac{k}{J}\rceil+2-\theta$, are non-overlapping. 
Among the non-overlapping local groups, $J-1$ groups have cardinality $\lfloor\frac{k}{J}\rfloor+1$, while the remaining have cardinality $\lceil\frac{k}{J}\rceil+1$.
Among the overlapping local groups, $\Big\lceil\frac{k}{J}\Big\rceil+1-\theta$ groups have cardinality $\lceil\frac{k}{J}\rceil+1$, while the remaining one has cardinality $\theta$.
This last overlapping local group shares exactly one VN with $(\Big\lceil\frac{k}{J}\Big\rceil+1-\theta)$ overlapping local groups, which is possible by (\ref{eq:cond_sec}).

Note that each VN has locality of $\lfloor\frac{k}{J}\rfloor$ or $\lceil\frac{k}{J}\rceil$, and the average locality is optimum by Theorem \ref{thm:genr_r_avg}. 
Also, it can be verified that every $J$ local CNs cover at least $J+k$ VNs, thus, by Corollary \ref{cor:d_cndn}, the minimum distance of $\mathcal{C}_{\overline{r}LRC_2}$ is at least $d$.

\subsection{The Third Class of $\overline{r}$-Optimal LRCs}\label{subsec:3rd}
Suppose the parameters $n$ and $k$ are such that $R>(1-\frac{1}{\sqrt{n}})^2$.
Then, the third class of our proposed $\overline{r}$-optimal LRCs ($\mathcal{C}_{\overline{r}LRC_3}$) achieving the bound in Theorem \ref{thm:avg_loc} is constructed through the following steps (Fig. \ref{fig:r_avg_prop}).\\ 
\textit{Step 1)} Obtain $\theta\in[0,d-2]$, denoted $\theta^*$, that minimizes (\ref{eq:thm}) in Theorem \ref{thm:avg_loc}.\\ \vspace{2mm}
\textit{Step 2)} Partition the set of $n$ VNs $\mathcal{Y}=\{y_1,\cdots,y_n\}$ into two subsets $\mathcal{Y}_A$ and $\mathcal{Y}_B$ with cardinality $n-{\theta^*}$ and $\theta^*$, respectively.\\
\textit{Step 3)} Partition $n-\theta^*$ VNs associated with $\mathcal{Y}_A$ into $J=n-k-d+2$ local groups as follows:
$J-a_{\theta^*}$ 
local groups with cardinality 
$\lfloor\frac{n-\theta^*}{J}\rfloor$;
and
	$a_{\theta^*}$ 
	local groups with cardinality 
	$\lceil\frac{n-\theta^*}{J}\rceil$, 
	where $a_{\theta^*}=n-\theta^*+J-J\Big\lceil \frac{n-\theta^*}{J}\Big\rceil$. 
	Therefore, there are 
	$(J-a_{\theta^*})\lfloor\frac{n-\theta^*}{J}\rfloor$ 
	VNs with locality
	$\Big\lfloor\frac{n-\theta^*}{J}\Big\rfloor-1$
	and  
	$a_{\theta^*}\lceil\frac{n-\theta^*}{J}\rceil$ 
	VNs with locality  
	$\Big\lceil\frac{n-\theta^*}{J}\Big\rceil-1$.\\
\textit{Step 4)} If $\theta^*\neq 0$, construct $(J+1)$-th local group with the following VNs:
	$i)$ $\lfloor\frac{n-{\theta^*}}{J}\rfloor-d+2$ VNs from each of local groups  
	$1$ to $J-a_{\theta^*}$;
	$ii)$ $\lceil\frac{n-\theta^*}{J}\rceil-d+2$ VNs from each of local groups 
	$J-a_{\theta^*}+1$ to $J$; and
	$iii)$ all the $\theta^*$ VNs associated with $\mathcal{Y}_B$.
	Hence, locality of the last local group is
	\begin{equation*}
	r_{J+1}=(J-a_{\theta^*})\Big(\Big\lfloor\frac{n-{\theta^*}}{J}\Big\rfloor-d+2\Big)+a_{\theta^*}\Big(\Big\lceil\frac{n-\theta^*}{J}\Big\rceil-d+2\Big)+\theta^*-1.
	\end{equation*}
	By manipulation, we have
	\begin{equation}\label{eq:r_3}
		r_{J+1} = n-J(d-2)-1.
	\end{equation}\\
\textit{Step 5)} Observe that there are $J+b_{\theta^*}$ local CNs and $n-k-(J+b_{\theta^*})$ global CNs, where $b_{\theta^*}=0$ if $\theta^*=0$ and $b_{\theta^*}=1$ otherwise.
	Connect each of the $n-k-J-b_{\theta^*}=d-2-b_{\theta^*}$ to all $n$ VNs.

\begin{remark}
	It can be verified that every $J$ local CNs cover at least $J+k$ VNs, thus, by Corollary \ref{cor:d_cndn}, the minimum distance of $\mathcal{C}_{\overline{r}LRC_3}$ is at least $d$.
\end{remark}

\begin{example}
In this example, we construct an $(n,k,d)=(8,4,4)$ LRC by following the aforementioned steps.
Here, $J=n-k-d+2=2$ and $\theta$ minimizing (\ref{eq:thm}) is $\theta^*=2$.
All the eight VNs are partitioned into two sets $\mathcal{Y}_A=\{y_1,\cdots,y_6\}$ and $\mathcal{Y}_B=\{y_7,y_8\}$.
Since $J\mid n-\theta^*$, there are $J=2$ local groups with cardinality $\frac{n-\theta^*}{J}=3$.
Since $\theta^*\neq 0$, the third local group has to be constructed by $\lceil\frac{n-\theta^*}{J}\rceil-d+2=\lceil\frac{8-2}{2}\rceil-4+2=1$ VN from each of the first two local groups and $\theta^*=2$ VNs associated with $\mathcal{Y}_B$.  
Let us form the third local group by VNs $y_3$, $y_6$, $y_7$, and $y_8$.
There is $n-k-J-b_{\theta^*}=8-4-2-1=1$ global check node which is connected to all $8$ VNs.
Fig. \ref{fig:1b} depicts Tanner graph associated with LRC of this example.
\end{example}


\begin{prop}\label{prop:ach_r_avg}
	$\mathcal{C}_{\overline{r}LRC_3}$	satisfies the bound on $\overline{r}$ in Theorem \ref{thm:avg_loc} with equality.
\end{prop}
\begin{IEEEproof}
	The first $n-\theta^*$ VNs of the code are partitioned into 
	$(J-a_{\theta^*})\lfloor\frac{n-{\theta^*}}{J}\rfloor$ 
	and 
	$a_{\theta^*}\lceil\frac{n-{\theta^*}}{J}\rceil$
	VNs with locality 
$\Big\lfloor\frac{n-{\theta^*}}{J}\Big\rfloor-1$
	and 
$\Big\lceil\frac{n-{\theta^*}}{J}\Big\rceil-1$
	, respectively.
	Also, locality of the last $\theta^*$ VNs is $n-J(d-2)-1$.
	Hence, for the average locality of $\mathcal{C}_{\overline{r}LRC_3}$, denoted $\overline{r}_{\mathcal{C}_3}$, we have
	\begin{align*} \label{eq:ravg0}
	\begin{split}
		n\overline{r}_{\mathcal{C}_3} &=
		(J-a_{\theta^*}) \Big\lfloor\frac{n-{\theta^*}}{J}\Big\rfloor \Big(\Big\lfloor\frac{n-{\theta^*}}{J}\Big\rfloor-1\Big)
		\\&
		+ a_{\theta^*}\Big\lceil\frac{n-{\theta^*}}{J}\Big\rceil\Big(\Big\lceil\frac{n-{\theta^*}}{J}\Big\rceil-1\Big)
		+\theta^*(n-J(d-2)-1) 
		\\&
		=(J-a_\theta^*)\Big\lfloor\frac{n-\theta^*}{J}\Big\rfloor^2+a_\theta^*\Big\lceil\frac{n-\theta^*}{J}\Big\rceil^2
		\\&
		-\Big(a_{\theta^*}\Big\lceil\frac{n-{\theta^*}}{J}\Big\rceil+ (J-a_{\theta^*})\Big\lfloor\frac{n-{\theta^*}}{J}\Big\rfloor\Big)
		\\&
		+\theta^*(n-J(d-2)-1).
	\end{split}
	\end{align*}
	By manipulation, we have
	 \[a_{\theta^*}\Big\lceil\frac{n-{\theta^*}}{J}\Big\rceil+ (J-a_{\theta^*})\Big\lfloor\frac{n-{\theta^*}}{J}\Big\rfloor=n-\theta^*.\]
	 Thus,
	\begin{equation}
	n\overline{r}_{\mathcal{C}_3}=(J-a_{\theta^*})\Big\lfloor\frac{n-{\theta^*}}{J}\Big\rfloor^2
	+a_{\theta^*}\Big\lceil\frac{n-\theta^*}{J}\Big\rceil^2+\theta^*(n-J(d-2))-n,
	\end{equation}
	which is equivalent to the minimum of $\overline{r}$ in Theorem \ref{thm:avg_loc}.
\end{IEEEproof}

%% file: ProposedLRC.tex
In this section, we show how our proposed approach can improve locality of the $(n,k,d,\overline{r})=(16,10,5,5)$ LRC used in Facebook HDFS-RAID \cite{XORingElephants}, denoted $\mathcal{C}_F$, without sacrificing its crucial parameters, namely the code rate $(R)$ and the code minimum distance $(d)$.
We show that by using our proposed LRC, the average locality of $\mathcal{C}_F$ is improved 22.5\%.

For $\mathcal{C}_F$, we have $R=0.625>(1-\frac{1}{\sqrt{n}})^2=0.56$.
Hence, the code construction described in Section \ref{subsec:3rd} can be used.
For an $(n,k,d)=(16,10,5)$ LRC, it is verified by Theorem \ref{thm:avg_loc} that $\overline{r}\geq 3.875$, which is obtained for $\theta=d-2=3$.
\begin{figure}[t!]
	\centering
	\includegraphics[width=0.5\textwidth, scale=0.6]{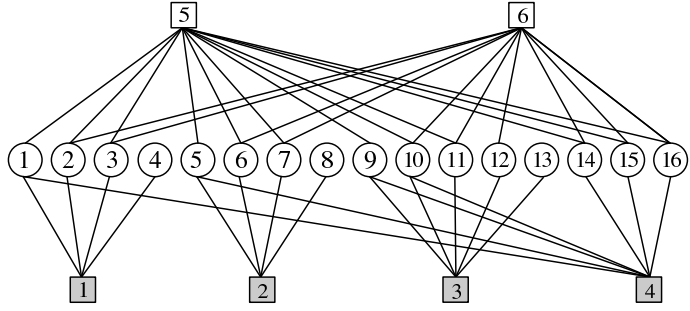}
	\caption{Tanner graph of our proposed $(n,k,d,\overline{r})=(16,10,5,3.875)$ LRC with four local CNs (gray squares) and two global CNs (white squares).}
	\label{fig:TannerC1}
\end{figure}
The Tanner graph of our proposed LRC, denoted $\mathcal{C}_0$, is presented in Fig \ref{fig:TannerC1}. 
By using the Tanner graph, the parity check matrix of $\mathcal{C}_0$, denoted $\mathbf{H}_0$, is constructed. 
The non-zero elements of $\mathbf{H}_0\in\mathbb{F}_{2^8}^{6\times 16}$ are numerically selected from the finite field $\mathbb{F}_{2^8}$ formed by the primitive polynomial $x^8+x^4+x^3+x^2+1$ such that every $d-1=4$ columns of $\mathbf{H}_0$ are linearly independent.
Then, by multiplying the inverse of a full-rank $6\times 6$ sub-matrix of $\mathbf{H}_0$ from the left by $\mathbf{H}_0$, the systematic form of $\mathbf{H}_0$ is obtained.
By using the obtained systematic form, the systematic form of the generator matrix of our proposed $(n,k,d,\overline{r})=(16,10,5,3.875)$ LRC, denoted $\mathbf{G}_0\in \mathbb{F}_{2^8}^{10\times 16}$, can be explicitly represented as follows.
\begin{equation*}\label{eq:G0}
\begin{split}
\mathbf{G}_0^T=~~~~~~~~~~~~~~~~~~~~~~~~~~~~~~~~~~~~~~~~~~~~~~~~~~~~~~~~~~~~~~~~~~~~~&\\
\left(\begin{array}{@{}*{16}{c}@{}}
0&0&1&0&0&0&0&0&0&0\\
0&0&0&1&0&0&0&0&0&0\\
0&0&0&0&0&0&1&0&0&0\\
0&0&0&0&0&0&0&0&0&1\\
35&134& 39& 29& 15&191&187&3&102& 38\\
34&135& 39& 29& 15&191&187&3&102& 38\\
234&137& 29&254&245&110&153&9&223&2\\
243&249& 60& 11& 59&234& 48& 37&217&104\\
25&112& 32&245&206&132&169& 44&6&106\\
0&0&0&1&1&1&1&0&0&0\\
1&0&0&0&0&0&0&0&0&0\\
0&1&0&0&0&0&0&0&0&0\\
0&0&0&0&1&0&0&0&0&0\\
0&0&0&0&0&1&0&0&0&0\\
0&0&0&0&0&0&0&1&0&0\\
0&0&0&0&0&0&0&0&1&0\\
\end{array} \right),~~~~~
\end{split}
\end{equation*}
where each element of $\mathbf{G}_0$, denoted $g_{i,j}$ for $i\in[1,16]$ and $j\in[1,10]$, is the decimal representation of a byte of the form $[a_7,\cdots,a_0]\in\mathbb{F}_2^{1\times 8}$, i.e. $g_{i,j}=\sum_{l=0}^{7}a_l2^l$ where $a_l\in\mathbb{F}_2$, $l\in[0,7]$.
For example, $35\equiv [0,0,1,0,0,0,1,1]$.

%% file: Conclusion.tex
The average locality of locally repairable codes (LRCs) is directly translated to the costly repair bandwidth of distributed storage systems.  
While the existing literature mainly focuses on the maximum locality of erasure codes used for distributed storage systems, the average locality $\overline{r}$ is also an important factor which directly affects costly repair bandwidth and disk I/O. 
In this paper, we used a novel approach to find a general lower bound on $\overline{r}$ of  erasure codes with arbitrary parameters. 
We also proposed LRCs that achieve the obtaining bounds, offering improvement on $\overline{r}$ over the existing locally reparable codes.

Generalizing our proposed lower bound on $\overline{r}$ for vector codes and erasure codes with multiple groups of locality as well as constructing explicit codes with small $\overline{r}$ can be of interest.



%% file: AppAvgPrf.tex
\begin{figure*}[t!]
	\centering
	\includegraphics[width=0.55\textwidth, scale=.9]{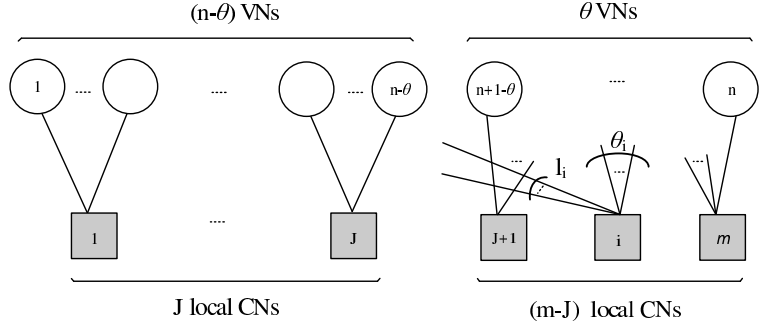}
	\caption{Locality Tanner graph of an $(n,k,d)$ erasure code with $m$ local CNs, where $m >J$. The first $J$ local CNs determine locality of the first $(n-\theta)$ VNs and the last $(m-J)$ local CNs determine locality of the last $\theta$ CNs, where $\theta\in[1,~d-2]$.}
	\label{fig:roptprov}
\end{figure*}

\begin{lem}\label{lem:eqvui_con}
For an $(n,k,d)$ code with $R> (1-\frac{1}{\sqrt{n}})^2$, we have
$d-3<\Big\lceil\frac{k}{n-k-d+2}\Big\rceil=\Big\lceil\frac{k}{J}\Big\rceil$.
\end{lem}
\begin{IEEEproof}
We have
\begin{equation*} 
\begin{split}
&~~~d-3<\Big\lceil\frac{k}{n-k-d+2}\Big\rceil 
	\\&
\Leftrightarrow (n-k-d+2)(d-3)< k 
	\\&
	\Leftrightarrow (d-2)^2-(n-k+1)(d-2) +n> 0 
	\\&
	\Leftrightarrow (\frac{n-k+1}{2})^2-(n-k+1)(\frac{n-k+1}{2}) +n > 0 
	\\&
	\Leftrightarrow 4n-(n-k+1)^2> 0
	\\&
	\Leftrightarrow\frac{k}{n}> (1-\frac{1}{\sqrt{n}})^2
\end{split}
\end{equation*}
where the forth inequality is true because the second degree equation $x^2-(n-k+1)x+n$ takes its minimum at $x=\frac{n-k+1}{2}$.
\end{IEEEproof}

\begin{lem}\label{lem:CardCoverage}
	For a linear block code with minimum distance $d$ and its corresponding locality Tanner graph $\mathcal{T}$, assume that \\
		1) total number of local CNs is greater than $J$, i.e. $m> J$;\\
		2) the first $J$ local CNs of $\mathcal{T}$ cover $(n-\theta)$ VNs, where $\theta\in[1,d-2]$; and\\
		3) among all $\theta$ VNs not covered by the first $J$ local CNs, only $\theta_i$ VNs are linked to $i$-th local CN, where $\theta_i\in [1,\theta]$ and $i\in[J+1,m]$ (Fig. \ref{fig:roptprov}).
	Then, each local group $\mathcal{Y}_l$ with cardinality more than $\theta_i$ has at least 
	$(|\mathcal{Y}_l|-\theta_i+\theta-d+2)$ VNs connected to local CN associated with local group $\mathcal{Y}_i$, where $l\in[1,J]$ and $i\in[J+1,m]$. 
\end{lem}
\begin{IEEEproof}
	Without loss of generality, assume that local group $\mathcal{Y}_J$ with $|\mathcal{Y}_J|>\theta_i$ has less than
	$(|\mathcal{Y}_J|-\theta_i+\theta-d+2)$ VNs in common with local group  $\mathcal{Y}_{J+1}$.  
	This implies that $J$ local CNs indexed by $\{1, 2, \cdots, J-1,J+1\}$ cover at most 
	\begin{equation*}
	\begin{split}
	&~~~\sum_{l=1}^{J-1}|\mathcal{Y}_l|+(|\mathcal{Y}_J|-\theta_i+\theta-d+2-1)+\theta_i
	\\&=\sum_{l=1}^{J}|\mathcal{Y}_l|+\theta-d+1=(n-\theta)+\theta-d+1 \\&= J+ k -1
	\end{split}
	\end{equation*}
	VNs which contradicts Lemma \ref{lem:14_4}.
\end{IEEEproof}

\begin{lem}\label{lem:dec_func}
	Consider function  $f(z_1,\cdots,z_m,m)=\sum_{i=1}^{m}{z_i}^2$, where $z_i$'s are positive non-zero integers for $i\in[1,m]$ which satisfy $\sum_{i=1}^{m}{z_i}=A$, and $A$ is a constant integer. 
	Assume that minimum of $f(z_1,\cdots,z_m,m)$ is attained when $z_i=z_i^*~\forall i$, i.e. 
	\[
	f_{\min}(m):=f(z_1^*,\cdots,z_m^*,m)=\underset{z_i}\min~f(z_1,\cdots,z_m,m).
	\]
	Then, $f_{\min}(m)$ is a decreasing function.
\end{lem}
\begin{IEEEproof}
	We need to show that $f_{\min}(m)\geq f_{\min}(m+1)$. 
	Assume that $f_{\min}(m)=f(w_1,\cdots,w_m,m)$ and $f_{\min}(m+1)=f(v_1,\cdots,v_{m+1},m+1)$ with $\sum_{i=1}^{m}w_i=\sum_{i=1}^{m+1}v_i=A$. We have, 
	\begin{equation*}
		\begin{split}
			f_{\min}(m) &= 
			\sum_{i=1}^{m}w_i^2= \sum_{i=1}^{m-1}w_i^2+(\frac{w_m}{2}+
			\frac{w_m}{2})^2
			\\&
			\geq\sum_{i=1}^{m-1}w_i^2+(\frac{w_m}{2})^2+
			(\frac{w_m}{2})^2 
			\\&
			\geq 	
			\sum_{i=1}^{m+1}v_i^2 =f_{\min}(m+1),
		\end{split}
	\end{equation*}	
	where the first inequality is by $(a+b)^2\geq a^2+b^2$.
\end{IEEEproof}

\vspace{.3cm}
\textit{Proof of Theorem \ref{thm:avg_loc}:}
By using Algorithm \ref{alg}, the set of $n$ VNs $\{y_1,\cdots,y_n\}$ is partitioned into $m$ local groups $\mathcal{Y}_1$ to $\mathcal{Y}_m$. 
Then, we calculate the average locality in terms of code's parameters for different values of $m\in[2,n-k]$.
More specifically, in order to evaluate $\overline{r}$, we assume that $m\in[2,J]$ and $m\in[J+1,n-k]$ for the first case and the second case, respectively.\\
case (i) $m\in [2,J]$: 
In this case, 
\begin{equation}\label{eq:caseA0}
n~\overline{r}=\sum_{i=1}^m|\mathcal{Y}_i|r_i\geq \sum_{i=1}^m|\mathcal{Y}_i|(|\mathcal{Y}_i|-1) =
\sum_{i=1}^m|\mathcal{Y}_i|^2-\sum_{i=1}^m|\mathcal{Y}_i| 
\end{equation}
Observe that $\sum_{i=1}^m |\mathcal{Y}_i|=n,~\forall m\in[2,J]$ because all the $m$ local groups have to cover all $n$ VNs.
By Lemma~\ref{lem:dec_func}, $\sum_{i=1}^{m}|\mathcal{Y}_i|^2$ is a decreasing function. Hence, its minimum is attained when $m=J$. Therefore, considering (\ref{eq:caseA0}) and Lemma~\ref{lem:OptSqrt}, we have
\begin{equation}\label{eq:caseA1}
\begin{split}
\overline{r}&\geq\frac{\sum_{i=1}^J|\mathcal{Y}_i|^2-n}{n}= 
\frac{\sum_{i=1}^J|\mathcal{Y}_i|^2}{n}-1 
\\&
\geq\frac{a_0\Big\lceil\frac{n}{J}\Big\rceil^2 +(J-a_0)\Big\lfloor\frac{n}{J}\Big\rfloor^2}{n}-1,~~~~~~~~~~~
\end{split}
\end{equation}
where $a_0=n+J-J\big\lceil\frac{n}{J}\big\rceil$.
Observe that replacing $\theta$ with zero represents (\ref{eq:caseA1}) in Theorem \ref{thm:avg_loc}. 
Now, we consider the second case where not all $n$ VNs are covered by the first $J$ CNs.

\vspace{.1cm}
case (ii)  $m\in [J+1,n-k]$: 
Considering Algorithm \ref{alg}, in this case, $\mathcal{U}_{J+1}=\mathcal{Y}_1\bigcup \cdots\bigcup \mathcal{Y}_J$. 
By Lemma~\ref{lem:14_4}, every $J$ CNs must cover at least $J+k$ VNs.
Hence, the first $J$ local CNs cover at least $J+k$ VNs, i.e. $|\mathcal{U}_{J+1}|=\sum_{i=1}^J|\mathcal{Y}_i|\in [J+k,n]$.
Observe that $\sum_{i=1}^J|\mathcal{Y}_i|=n$ is equivalent to the first case with $m=J$. 
In the following, we consider the remaining cases where $J$ local CNs cover $\sum_{i=1}^J|\mathcal{Y}_i|\in[J+k,n-1]$ VNs. 
Equivalently, the case where $\theta :=|\overline{\mathcal{U}}_{J+1}|=n-|\mathcal{U}_{J+1}|\in [1,d-2]$ VNs are not covered by the first $J$ local CNs. 
\balance

 

Observe that total number of local CNs is $m\in[J+1,n-k]$ among which the first $J$ local CNs determine locality of the first $n-\theta\in[n-(d-2),n-1]$ VNs and the remaining $(m-J)\in[1,d-2]$ local CNs determine locality of $\theta\in[1,d-2]$ VNs not covered by the first $J$ local CNs (Fig.~\ref{fig:roptprov}). 
We have
\begin{equation}\label{eq:rNonover}
\begin{split}
n\overline{r}&=\sum_{i=1}^{m}|\mathcal{Y}_i|r_i=
\sum_{i=1}^J|\mathcal{Y}_i|r_i+\sum_{i=J+1}^{m}|\mathcal{Y}_i|r_i\\&\geq
\sum_{i=1}^J|\mathcal{Y}_i|(|\mathcal{Y}_i|-1)+\sum_{i=J+1}^{m}|\mathcal{Y}_i|r_i\\&\geq
\sum_{i=1}^J{|\mathcal{Y}_i|}^2-\sum_{i=1}^J{|\mathcal{Y}_i|}+\theta \min\limits_{i\in[J+1, m]}r_i,
\end{split}
\end{equation}
where $\sum_{i=1}^J{|\mathcal{Y}_i|}=n-\theta$. Note that in (\ref{eq:rNonover}), the last inequality holds because locality of $\theta$ VNs not covered by the first $J$ local CNs is determined by the last $m-J$ local CNs. 

Now, we obtain $\min\limits_{i\in[J+1, m]}r_i$. 
Assume that among all $\theta$ VNs not covered by local CNs 1 to $J$, only $\theta_i$ VNs are linked to $i$-th local CN, where $\theta_i\in [1,\theta]$ and $i\in[J+1,m]$.
Observe that in order to satisfy Lemma~\ref{lem:14_4}, $i$-th CN must have some VNs, denoted $l_i$, in common with local groups $1$ to $J$ (Fig. ~\ref{fig:roptprov}).
Hence, locality associated with $i$-th local CNs for $i\in[J+1,m]$ is obtained as follows 
\begin{equation}\label{eq:r_i}
r_i = l_i+ \theta_i -1,~~~~i\in[J+1,m]. 
\end{equation}
Assume that among local CNs 1 to $J$, $\delta_i$ local CNs have cardinality more than $\theta_i$ and the rest $(J-\delta_i)$ local CNs have cardinality less than or equal to $\theta_i$, with $\delta_i\in [1,J-1]$. In other words, 
\begin{equation}\label{eq:deltaPrime}
\left\{ 
\begin{array}{l l}
|\mathcal{Y}_l|\leq\theta_i , & \quad l\in [1,J-\delta_i]\\
|\mathcal{Y}_l|> \theta_i , & \quad l\in[J+1-\delta_i,J] 
\end{array}
\right .
\end{equation}
By Lemma~\ref{lem:CardCoverage}, $l$-th local CN with $|\mathcal{Y}_l|>\theta_i$ must have at least $(|\mathcal{Y}_l|-\theta_i+\theta-d+2)$ VNs in common with $i$-th local CN, where $l\in[J+1-\delta_i,J]$ and $i\in[J+1,m]$.
Hence, we have 
\begin{equation}\label{eq:l_i}
l_i\geq\sum_{l=J+1-\delta_i}^J(|\mathcal{Y}_l|-\theta_i+\theta-d+2)=\sum_{l=J+1-\delta_i}^J|\mathcal{Y}_l|-\delta_i(\theta_i-\theta+d-2).
\end{equation}
Note that the first $J$ CNs cover $(n-\theta)$ VNs, that is to say $\sum_{l=1}^J|\mathcal{Y}_l|=n-\theta$. 
Thus, from (\ref{eq:deltaPrime}), we have
\begin{equation}\label{eq:NmTg}
\sum_{l=J+1-\delta_i}^{J}|\mathcal{Y}_l|=(n-\theta)-\sum_{l=1}^{J-\delta_i}|\mathcal{Y}_l|.
\end{equation}
Considering (\ref{eq:r_i}) to (\ref{eq:NmTg}) and noting that $\max\limits_{l\in[1, J-\delta_i]}|\mathcal{Y}_l|=\theta_i$, $\delta_i\leq J-1$, and $\theta_i\leq\theta\leq d-2<\Big\lceil\frac{k}{J}\Big\rceil+1$, 
we have the following chain of inequalities
\begin{equation*}\label{eq:rifinal}
\begin{split}
r_i&\geq n-\theta -\sum_{l=1}^{J-\delta_i}|\mathcal{Y}_l|-\delta_i\theta_i+\delta_i\theta-(d-2)\delta_i+\theta_i-1
\\&\geq n-\theta-(J-\delta_i)\theta_i-\delta_i\theta_i+\delta_i\theta-(d-2)\delta_i+\theta_i-1
\\&=n-\theta-(J-1)\theta_i+\delta_i\theta-(d-2)\delta_i-1\\&\geq
n-\theta-(J-1)\theta+\delta_i\theta-(d-2)\delta_i-1\\&=
n-(J-\delta_i)\theta-(d-2)\delta_i-1\\&\geq
n-(J-\delta_i)(d-2)-(d-2)\delta_i-1\\&=n-J(d-2)-1.
\end{split}
\end{equation*}
In other words, 
\begin{equation}\label{eq:min_ri}
\min\limits_{i\in[J+1, m]}r_i = n-J(d-2)-1.
\end{equation}
By replacing (\ref{eq:min_ri}) in (\ref{eq:rNonover}), we have:
\begin{equation}\label{eq:r_avg0}
n~\overline{r} \geq \sum_{i=1}^J{|\mathcal{Y}_i|}^2-\sum_{i=1}^J{|\mathcal{Y}_i|}+\theta(n-J(d-2)-1),~\theta\in[1,J]
\end{equation}
where $\sum_{i=1}^J|\mathcal{Y}_i|=n-\theta$. Hence, for $\theta\in[1,d-2]$
\begin{equation}\label{eq:r_avg1}
\begin{split}
n\overline{r} &\geq \sum_{i=1}^J{|\mathcal{Y}_i|}^2-(n-\theta)+\theta(n-J(d-2)-1)
\\&=\sum_{i=1}^J{|\mathcal{Y}_i|}^2-J\theta(d-2)+n\theta-n.
\end{split}
\end{equation}
By Lemma~\ref{lem:OptSqrt}, we have
	\begin{equation}\label{eq:tet_1J}
	\small
	\overline{r}\geq 
	\frac{
		\underset{\theta\in[1,d-2]}\min	
		\Bigg\{ 
		(J-a_\theta)\Big\lfloor\frac{n-\theta}{J}\Big\rfloor^2
		+a_\theta\Big\lceil\frac{n-\theta}{J}\Big\rceil^2
		+(n-dJ+2J)\theta
		\Bigg\}
	}{n}
	-1,
	\end{equation} 
Combining (\ref{eq:caseA1}) and (\ref{eq:tet_1J}) concludes the proof.

\vspace{.25cm}
%
%